\documentclass[conference]{IEEEtran}

\usepackage{hyperref}
\usepackage{cite}
\usepackage{amsmath,amssymb,amsfonts}
\usepackage{algorithmic}
\usepackage{graphicx}
\usepackage{textcomp}
\usepackage{xcolor}
\def\BibTeX{{\rm B\kern-.05em{\sc i\kern-.025em b}\kern-.08em
    T\kern-.1667em\lower.7ex\hbox{E}\kern-.125emX}}

\makeatletter
\def\endthebibliography{%
  \def\@noitemerr{\@latex@warning{Empty `thebibliography' environment}}%
  \endlist
}
\makeatother

\begin{document}

\title{Sludge for Good: Slowing and Imposing Costs on Cyber Attackers}

\author{
\IEEEauthorblockN{Josiah Dykstra}
\IEEEauthorblockA{\textit{Cybersecurity Collaboration Center} \\
\textit{National Security Agency}\\
Ft. George G. Meade, USA \\
josiah.dykstra@cyber.nsa.gov}
\and
\IEEEauthorblockN{Kelly Shortridge}
\IEEEauthorblockA{
\textit{Fastly, Inc.}\\
San Francisco, USA \\
kelly@shortridge.io}
\and
\IEEEauthorblockN{Jamie Met}
\IEEEauthorblockA{\
\textit{National Security Agency}\\
Ft. George G. Meade, USA \\
jlmet@nsa.gov}
\and
\IEEEauthorblockN{Douglas Hough}
\IEEEauthorblockA{
\textit{Johns Hopkins}
\textit{Bloomberg School of Public Health}\\
Baltimore, USA \\
douglas.hough@jhu.edu }
}

\maketitle

\begin{abstract}
Choice architecture describes the design by which choices are presented to people. Nudges are an aspect intended to make ``good'' outcomes easy, such as using password meters to encourage strong passwords. Sludge, on the contrary, is friction that raises the transaction cost and is often seen as a negative to users. Turning this concept around, we propose applying sludge for positive cybersecurity outcomes by using it offensively to consume attackers’ time and other resources. 

To date, most cyber defenses have been designed to be optimally strong and effective and prohibit or eliminate attackers as quickly as possible. Our complimentary approach is to also deploy defenses that seek to maximize the consumption of the attackers’ time and other resources while causing as little damage as possible to the victim. This is consistent with zero trust and similar mindsets which assume breach. The Sludge Strategy introduces cost-imposing cyber defense by strategically deploying friction for attackers before, during, and after an attack using deception and authentic design features. We present the characteristics of effective sludge, and show a continuum from light to heavy sludge. We describe the quantitative and qualitative costs to attackers and offer practical considerations for deploying sludge in practice. Finally, we examine real-world examples of U.S. government operations to frustrate and impose cost on cyber adversaries.

\end{abstract}

\begin{IEEEkeywords}
sludge, nudge, cybersecurity, choice architecture, deception
\end{IEEEkeywords}

\section{Introduction}

In their 2009 book, Richard Thaler and Cass Sunstein introduced the concept of ``nudge,'' which they defined as an intervention that ``alters people’s behavior in a predictable way without forbidding any options or significantly changing their economic incentives.''~\cite{thaler2009nudge} Thaler admonishes policy makers to ``nudge for good.'' This concept has been remarkably popular, and has been applied in such diverse areas as registration for organ donation, contributions to retirement accounts, and consumption of healthy food. Recently, proposals have been made to apply nudges to cybersecurity. For instance, password meters encourage users to select strong passwords. 

In a later edition of their book (and in a separate book by Sunstein~\cite{sunstein2021sludge}), Thaler and Sunstein introduced a contrary concept: sludge, defined as ``any aspect of choice architecture consisting of friction that makes it harder for people to obtain an outcome that will make them better off'' ~\cite{thaler2021nudge}. Sludge takes many forms including excessive wait times, unjustified overhead, and dreary requirements. It is friction that raises the transaction cost and is often seen as a negative for people with benevolent goals. Classic examples of sludge are: streaming services that offer free or low-cost subscriptions that automatically convert to much higher fees at the end of the trial period, and which require subscribers to go through multiple steps to cancel; and mail-in vouchers that require customers to provide extensive documentation (much of which the customers rarely retain). Thaler also cites phishing warnings as an example of sludge. When a user predominately receives emails from external senders, a warning that the email is from an external sender becomes nothing more than visual clutter that wastes time and attention---unnecessary friction that distracts from the goal of consuming and responding to email, without much benefit to the purported benefit of greater security.

The kinds of sludge considered by Thaler and Sunstein have focused on ``bad'' sludge. In this paper, we propose ``sludge for good'' in cybersecurity. That is, we discuss sludges designed to slow down cyber attackers. 
If it is possible to influence attackers’ choices and slow them down, this can help deter, prevent, or raise the cost of achieving their goals. This is not unprecedented. Strategically slowing an advancing adversary is an ancient goal in physical world conflict where castle moats, buried mines, and road barricades have been used.

Slowing attackers raises the opportunity cost for attackers. Conducting a successful attack is not free for attackers. At the very least they require time to choose targets, conduct the attack, and determine when to abandon the attack. From the attacker’s perspective, dealing with sludge costs them money, time, and toil. Attackers also give up any gains they might have acquired if they had chosen other attacks or other victims. 

Deliberately influencing attackers’ decisions and behavior is an emerging idea. U.S. Cyber Command talks about ``imposing cost'' on adversaries~\cite{nakasone_2020}. This view implies offensive consequences and retaliation for bad behavior but could be expanded to include defensive tactics that make it more difficult and expensive for adversaries to conduct cyber attacks. Attackers expect their campaigns to generate a positive return on investment and will consider what costs 
are appropriate relative to the benefits of achieving a given goal. Budgets and resources are finite so this constrains the spectrum of actions an attacker can execute in a given campaign. Certain kinds of attacks, such as supply chain backdoors for example, generally require the skills and resources of a nation state.

To date, most cyber defenses have been designed to be optimally strong and effective and prohibit or eliminate attackers as quickly as possible. We propose a complimentary approach which is to also deploy defenses that seek to maximize the consumption of the attackers’ time and other resources while causing as little damage as possible to the victim. This is consistent with the ``assume compromise'' mindset whereby defenders treat all systems as not secure and already compromised. Slowing attackers is also complementary to a zero trust architecture where no system, user, or data are explicitly trusted. To understand and effectively impose cost by slowing attackers, new theory and doctrine are needed to create a common frame of reference. We present the characteristics of good sludge and provide a continuum from light to heavy options. We conclude with some recommendations for practitioners.

\section{Nudges in Cybersecurity}

Thaler and Sunstein use the phrase ``choice architecture'' to describe the design in which choices are presented to people. These choices are made easier by nudges and more difficult by sludge. To understand choice architecture, it is necessary to examine both nudge and sludge. Nudges gently steer people in a direction that increases welfare, including cybersecurity, and are commonly intended to make good outcomes easy. Traditionally, nudges have been used to encourage well-intentioned users to behave in a way that they are better off for doing so. These choices are not guaranteed, but research shows that they are selected more often.

Researchers in one study conducted a literature review of 71 papers on technology-mediated nudges~\cite{caraban201923}. They identified 23 distinct mechanisms of nudging which they grouped into six categories: facilitate, confront, deceive, social influence, fear, and reinforce. Only 9\% of the papers reviewed were related to security and privacy, and none used deception to promote a particular outcome for either legitimate users or attackers.

Prior work has explored nudges for cybersecurity in depth, particularly for end users. Zimmermann and Renaud established that nudges in cybersecurity have four attributes: ``predictability of the influence and outcome, involvement of automatic cognitive processes, equality of choice costs, and retention of all pre-nudge choices''~\cite{zimmermann2021nudge}. They applied these principles to decisions such as choosing more secure public WiFi networks and encrypting a smartphone. Peer et al. found that nudges aimed at average users were less effective than nudges that were personalized by decision-making style~\cite{peer2020nudge}. Based on their calculations, the median time to crack passwords was 4.2 times longer for personalized password nudges than for random assignments or no nudge at all.

Nudges have also been studied to help software developers make secure choices. One recent study attempted to nudge developers into using safe code snippets on the web rather than unsafe code~\cite{fischer2022nudging}. Among 218 participants, those receiving nudges produced more functional and secure code in comparison to the control group.

Because they are optional, nudges are not always successful in achieving a specific outcome. Twitter, for example, encourages two-factor authentication (2FA) but reports that only 2.5\% of users have it enabled~\cite{twitter_2021}. In a recent large-scale in-the-wild study of 2FA adoption messages on Facebook, researchers found that messaging and design strategies can increase adoption, especially when incorporating personalized prompts~\cite{golla2021driving}. Nevertheless, 2FA remains an opt-in choice on most platforms. When prompts are too aggressive and persistent, nudges produce negative by-products. Users of Apple products are continually reminded to enable 2FA even though it is optional, resulting in user irritation~\cite{apple_2017}.

There is some evidence that defenders can nudge adversaries in a manner that is beneficial for the defender yet detrimental for the attacker. Some researchers have considered a defensive goal of delaying attackers, rather than denying them, using decoy systems with packet delays~\cite{landsborough2021towards}. In a laboratory test, the delay tactic increased the average run time from 243.0 seconds (control) to 687.62 seconds and decreased attacker success, while a denial tactic occupied less time and produced lower attacker success. ``This technique,'' they write, ``provides the appearance of poor network performance which can nudge an attacker into moving to a new target---useful if the attacker’s original target was a valuable or vulnerable system.''
Others have examined oppositional human factors as a domain of deception intended to nudge attackers into negative affective states~\cite{ferguson2021oppositional}. In particular, they found in one penetration experiment with professional red teamers where deception was used that participants exhibited confusion, self-doubt, surprise, and frustration. Other work in deception has continued to examine not only discovering adversary activity or deflecting them, but also in depleting their resources. For example, another study explored deception for malware using honey files and honey credentials~\cite{sajid2021soda}. The researchers measured 13-15\% overhead to the system owner of orchestrating two depletion deception techniques but did not attempt to measure the cost borne by the attacker. 

In 2017, Shortridge proposed a new defensive framework that leverages ``learning exploitation,'' raising the cost of attack by destabilizing attackers’ ability to learn ~\cite{shortridge_2017}. Within learning exploitation, this approach included sludge-like interventions that make it harder for attackers to learn information about target systems and that introduce unreliability into attack operations. Bogus credentials, for instance, waste attacker time and attention, slowing down their operations. This work also proposed introducing strategic non-determinism into systems to raise the cost of attack during the reconnaissance phase. For example, a defender could make normal endpoints appear like a different malware analysis sandbox upon each startup by adding ``hollow but sketchy-looking artifacts'' such as debuggers and virtualization libraries.

Two practitioners have reinvigorated the design of deception environments to add ``anticipatory mechanisms that impede the success of [attackers’] operations''~\cite{shortridge2021lamboozling}. ``Understanding how attackers make decisions allows software engineers to exploit the attackers' brains for improved resilience,'' they write. The goal of their proposed deception environments is to ``disrupt attackers’ abilities to learn and make decisions,'' introducing friction into attack operations that slows down attackers while also allowing defenders to collect information about attacker actions that can aid in defense. 

Nudges play a role in the choice architecture of cyber defense in helping make a decision. A defender can influence an attacker's behavior to a destination or activity that controls the damage. Nudges alone do not account for the experience that follows the commitment to a choice. What happens after an attacker chooses to explore a decoy system, for example, is when the cost of the choice is paid.

\section{Sludge Strategy for Slowing Attackers}
We propose a Sludge Strategy for cyber defense that prioritizes investments into techniques, tools, and technologies that add friction into attacker workflows and raise the cost of conducting operations. Defensive choice architects can leverage all forms of cost when designing sludge interventions against attackers. 
To aid in the adoption of weaponized choice architecture, we developed a strategy to help defenders understand the spectrum of potential sludge interventions to deploy against attackers. Table~\ref{table1} illustrates how selected defensive techniques offer degrees of sludge. Each row contains a defensive technique and describes cost(s) imposed on an attacker. Light sludge describes effects that produce low friction for attackers and heavy sludge produces high friction.
The heavier the sludge, the harder it is for attackers to trudge through it. 

\begin{table*}[htbp]
\caption{Examples of light to heavy sludge in cybersecurity and impact on attackers.}
\begin{center}
\begin{tabular}{p{0.25\textwidth}|c|c|c|c}
& \textbf{Type$^{1}$} & \textbf{Light Sludge$^{2}$} &  \textbf{Medium Sludge$^{2}$}& \textbf{Heavy Sludge$^{2}$}\\
\hline

Login Banners           & A     & M     &           &       \\
\hline

Authentication          & A     & I, T  &           &       \\
\hline

Decompression Bombs     & D     &       & M, T      &       \\
\hline

Network Throttling      & D     &       & P, T      &       \\
\hline

Perception of Deception & D     &       & P, T      &       \\
\hline

Immutable and ephemeral infrastructure & D  &   & I, M, P, T & \\
\hline

Deception Environment   & D     &       &           & I, T, P\\
\hline

Outing Tools / Infrastructure & A & & & M, P\\
\hline

Public Attribution/ Outing Attackers & A & & & P\\
\hline

Sanctions               & A     &       &           & M, P\\
\hline

\multicolumn{5}{l}{\footnotesize{$^1$ A: Authentic Design Feature, D: Deception}}\\
\multicolumn{5}{l}{\footnotesize{$^2$ I: Information Cost, M: Monetary Cost, P: Psychological Cost, T: Time Cost}}\\
\end{tabular}
\label{table1}
\end{center}
\end{table*}

\subsection{ Quantitative and Qualitative Costs}
The costs borne by attackers are both quantitative and qualitative. This section presents relevant aspects of each type and some illustrative examples.

\textbf{Quantitative Costs.} 
Most attackers must spend money to develop or purchase tools and technologies (including exploits and infrastructure) and pay for talent on their teams, which can be highly specialized (exploit developers, operators, etc.). The ransomware operator Trickbot purportedly invested more than \$20 million into ``infrastructure and growth of their organization'' in 2021 alone, including investment in technology, human capital, communications, software development, and extortion activities~\cite{burgess202_inside}. The LAPSUS\$ group, an amateur cybercriminal organization, attempted to recruit insiders by offering \$20,000 per week to employees willing to hand over their remote access credentials. Nation states may apply even more investment than cybercriminal organizations, especially in specialized skill sets. By one estimate, the Stuxnet attack costs the offense \$300 million~\cite{slayton2016cyber}. Attack operations require careful budgeting across a variety of activities.

Attackers must consume time to conduct their operations, which is in limited supply. Each unit of time they expend on one activity cannot be spent on another. Cybercriminals may have target revenue goals within a given quarter or year and nation state attackers may have mission goals within a specific time frame, too. If enough time passes without successful compromise, the business or mission will suffer---or be abandoned in favor of a new endeavor. Time costs can be qualitative as well. Even cybercriminals are conscious of burnout if they continuously work long hours~\cite{burgess_2022}.

\textbf{Qualitative Costs.} 
The qualitative costs of sludge should not be underestimated and can include information, psychological, and reputational costs~\cite{sunstein2020sludge}. 

Attackers require information about targets and their systems to be successful and collecting information can be a form of sludge if it takes effort to acquire. In order to send a spearphishing email an attacker must know the victim's address. The more difficult it is to find, process, or evaluate this information the more friction is imposed. Conversely, attackers are also susceptible to information overload. Excess information impedes the decision-making process, resulting in a poor decision or decision paralysis.

Attackers, being human, experience negative psychological impacts such as feeling frustrated, dismayed, ashamed, inferior, confused, helpless, stressed, worried, or discouraged~\cite{ferguson2021oppositional}. Not only can sludge induce these effects, but the effects are sludge to attackers because they impose undesirable friction to achieving their objectives. Some validated scales do exist to measure psychological impact, but they require interaction with the subject which is seldom available from cyber attackers. The Cyber Operations Stress Survey, for instance, uses self-reported measures of fatigue and cognitive workload~\cite{dykstra2018cyber}. However, the developers of this instrument found that cyber operations longer than five hours had significant effects on fatigue and frustration and nearly all cognitive workload factors. New approaches will be required to passively infer the effectiveness of imposing psychological cost on attackers.

There is insufficient insight and scientific study about the possible reputational costs to attackers. In theory, successful attacks could raise threat actors' reputation and unsuccessful or leaked activity could lower reputation. In public relations, metrics include sentiment analysis, stakeholder surveys, and opinion polls; these do not appear to be used to measure threat actor reputation today.

\subsection{Types of Sludge}
Sludge can be implemented using both authentic design features and deception. Authentic features are native to products and utilized by end users and administrators to protect their own use of a system. Legitimate system users, for instance, have passwords to authenticate themselves that can also impose friction on an attacker who wishes to access a target. Deception by creating fictitious data---including fake accounts, database records, files, and systems---can also impose friction since accessing that data can alert an administrator or invoke a defensive response. The names of these fictitious items traditionally carry the prefix ``honey-,'' as in honeypot or honeytoken, named for their ability to attract and trap attackers.

\textbf{Authentic Design Features.}
System owners commonly employ warnings, notices, and other banners before users log in to notify users of acceptable use. The U.S. Department of Defense, for instance, requires a standard mandatory notice and consent on all systems~\cite{stig_2015}. Users must acknowledge such messages before gaining access. These messages may also seek to deter unauthorized users as a ``no trespassing'' sign for fear of monitoring and prosecution. It is unlikely that banners impose friction on attackers, but their existence does establish terms of unauthorized access.

Authentication can be an example of sludge. Passwords, pins, biometrics, and other authentication mechanisms are an intentional barrier to keeping unauthorized users out of accounts and services. They are prominent for legitimate users and a common irritation, despite significant research and engineering in usability. In the best case, authentication is a minor inconvenience to users and a high cost to an adversary. Recently, vendors including Microsoft and Apple have shown support for passwordless logins for added security and convenience. Passwords appear poised to be replaced with security keys and multifactor authentication over time. Other examples of authentication sludge include login push notifications, which validate login requests by notifying an associated mobile device, and periodic key rotation, a best practice to limit the number of messages encrypted with the same key which helps prevent cryptanalysis attacks.

Defenders have a distinct advantage over attackers in the ability to control system accessibility, speed, and responsiveness. For example, the owner of a system can limit the number of login attempts before forcing a discretionary account lockout period. Network throttling allows a network owner to slow down a suspected attacker or aggressive user by limiting the communication speed of data flowing in or out. Rate limiting allows application owners to slow down automated attacks against APIs by constraining the number of requests that can reach the application server. Sessions timeouts are another example of limiting accessibility and imposing sludge if attackers must deal with losing access to systems that disconnects or log out after a set amount of time. Access sludge can also be imposed if systems only accept connections for predefined, trusted IP addresses.

On a binary level, an authentic feature that produces sludge for attackers is code obfuscation. This approach increases the time, psychological, and information costs necessary for attackers to discover vulnerabilities. Software developers can take steps to make reverse engineering and vulnerability discovery more difficult and time consuming, such as symbol stripping and anti-debugging techniques~\cite{votipka2020observational}. To our knowledge, the costs imposed by these techniques has not been measured. In recent research, a new mitigation showed promise in injecting delays into the execution of illicit cryptomining on continuous integration (CI) platforms~\cite{li2022robbery}. Their evaluation also showed that this rendered the attack unprofitable and with only small impacts on legitimate CI jobs.

Defenders can design their own infrastructure to be immutable and ephemeral, as is becoming an emerging trend in private sector defense through the practice of Security Chaos Engineering. Immutability means that once infrastructure is deployed, it cannot be changed. Attackers frequently take advantage of secure shell (SSH) access in servers; but an immutable server can have SSH access disabled by default, given no changes are allowed, cutting off that attack path. Ephemerality means that infrastructure lives only for a short period of time, usually the duration of executing a task, before terminating. A serverless function, for instance, may only allow attackers mere seconds of access before being killed automatically, making persistence extremely difficult. There may be some impact to end users, specifically software engineers who interact with this infrastructure, in that they may need to alter workflows and update design accordingly (such as finding other ways to troubleshoot problems in production if debugging is not allowed). However, this friction is asymmetric in its impact because it much more drastically changes how attackers engage with the target system than it does for software engineers.

In addition to technical capabilities, an authentic feature of cybersecurity is attribution. This response involves a government or private entity publicly naming an actor or nation-state responsible for particular cyber activity. Attribution is friction for adversaries for several reasons. Attribution draws attention to the activity which prepares and informs a broader community to prepare, detect, and evict similar activity in their environments. Few comprehensive analyses exist which measure the costs borne by attackers because of attribution. However, attribution can also lead to tangible economic sanctions and criminal indictments.

Among the heaviest sludge and most friction for attackers are political, economic, and criminal responses, such as sanctions and indictments. While the costs are more transparent, the impact in cybersecurity may be limited~\cite{romanosky2021private}. These options are available only to nation states. Operational examples are presented in Section~\ref{sec:SludgeInOperation}.

\textbf{Deception.}
System owners also have the upper hand in ground truth. System owners possess information about their systems that attackers must endeavor to acquire, reflecting an information asymmetry that can be exploited. Attackers must expend effort both in acquiring information and determining its relevance to their operations. System owners can leverage deception to lead attackers to operate based on false assumptions---resulting in wasted financial, time, and cognitive resources---or to foment fear, uncertainty, and doubt in attackers, who must then expend more resources attempting to differentiate the real from the mirage. 

The perception of deception is an effective form of sludge. In a study of 130 professional red teamers, psychological deception appeared to be effective even if the attacker merely believed it may be in use~\cite{ferguson2021examining}. This approach provides an unusually high return on investment for defenders.

Decompression bombs are a form of deception sludge that impose friction on attackers after they have stolen data. This technique works by enticing an attacker to steal a seemingly valuable file with a specially-created decoy that requires an excessive amount of time, disk space, or memory for the attacker to decompress. The approach increases the cost imposed compared with traditional non-compressed decoy files.

Honeypots to deceive and distract adversaries have been applied since the 1980s. Many honeypots are implemented as standalone systems to distract attackers from production systems. Few studies have examined the prevalence and success of honeypots on a large scale. In one study, researchers discovered over 19,000 Internet-facing honeypots in 637 autonomous systems~\cite{morishita2019detect}. This could be an underestimate but given that there are 100,000 autonomous systems on the Internet it likely shows that honeypots are rare. While a honeypot is a lure for attackers to a decoy system, tarpits further explicitly aim to slow the attacker. One Internet-wide scan found 215,000 IP addresses in 107 networks among 77 autonomous systems exhibiting tarpit-like behavior~\cite{alt2014uncovering}.

Related to honeypots, honey-patches are a technique where a defender patches a known vulnerability but adds functionality to mislead attackers into believing that a failed exploitation attempt was successful~\cite{araujo2014patches}. The attacker interacts with a decoy environment that consumes time.

\subsection{Practical Considerations}
Creating and deploying sludge will depend on business decisions including financial implementation costs, impact to legitimate users, and degree of transparency of the sludge to attackers. Sludge can be a by-product of some authentic features such as authentication. However, there are financial and time overhead costs to defenders in the maintenance and sustainability of deception sludge. Minimizing the friction imposed on defense remains an area of active research~\cite{sajid2021soda}. The impact to legitimate users may be inversely proportional to the amount of friction imposed on attackers. Logon banners, for example, are significantly more sludge to users than attackers while deception is significantly more sludge to attackers while low impact to users. Transparency of sludge will likely impact the degree to which it is effective. In the physical world, visible security such as cameras is known to be a deterrent but this may not be universally true for cyber deception.

There are circumstances when sludge is not desirable or not achievable. Sludge should not be used when it costs more to employ than the value it produces. Network owners must also evaluate legal issues including those related to liability, entrapment, and privacy.

\textbf{Relationship to Deterrence.} 
Sludge, and other approaches to impose cost by influencing attackers' decisions, will have an effect on nascent efforts to apply deterrence theory into effective cyber applications. A significant change in cyber deterrence is the need to project security and imply a negative cost should the attacker attempt to compromise a target, all without revealing capability. While no single approach which affects the attacker's decision calculus may alone be an effective deterrent, the combined effects of the choice architecture determine the  defenders' overall deterrence posture. Nudges that promote safer behaviors by end users increase sludge and offer fewer opportunities for attackers to gain access to target systems. Sludge also serves as a distinct deterrence toolkit by encouraging attackers to look elsewhere, sowing doubt on the validity of their access, and providing false appearances of successful compromise.

\textbf{Measuring Success.} 
Measures of effectiveness are an important aspect for evaluation. Some types of sludge produce data that are visible and measurable. For instance, a defender-controlled honeypot can be monitored by its owner. System owners also have insight into who triggers network throttling, when it occurs, and for how long it took effect. On the other hand, other measures of cost---such as attacker frustration---are more difficult to passively observe and measure. One way to measure success is by comparing the impacts on two networks attacked by the same threat actor where one applies sludge and one does not.

Traditional measures in cybersecurity incident response such as mean-time-to-discovery or mean-time-to-remediation are imprecise for sludge. Instead, new measurements will be necessary. These must be integrated and considered in consort with other defenses to form a composite picture of overall organizational cybersecurity.
\section{Cyber Sludge in Operations}
\label{sec:SludgeInOperation}

Three events over the past three years have illustrated actions consistent with slowing cyber attackers using sludge: defense of the 2020 U.S. elections, counter-ransomware efforts, and responses to Russia's invasion of Ukraine. In this section we describe how these examples demonstrate and achieve sludge-like impacts.

Sludge was not inevitable for any of these events. The cybersecurity community in the public and private sectors could have exclusively pursued zero tolerance, complete elimination of the problems using technical and non-technical solutions. Instead, these examples offer support that slowing adversaries was a component of the strategy. 

With regards to countering adversarial cyber activities, the United States has released public comments primarily around sanctions and outing attacker behavior rather than about specific technical details of online operations. This approach is consistent with the precedent of protecting sources and methods. Revealing operational details such as the use of network throttling, for example, may give attackers knowledge to potentially detect and avoid the sludge. One result is that it is easier for non-government observers to see qualitative costs than precise quantitative costs. Public discussion of sludge-producing authentic design features are more common than discussions about deception.

The U.S. has spoken generally about its desire to slow and disrupt attackers. In a recent interview, the National Security Agency's (NSA) Chief of Adversary Defeat said that 
``What we really want to try to do is aggravate, disrupt the adversary so they can't do the things they want to do -- doesn't mean we're going to stop them, right? These are persistent adversaries, but to make it harder, to make them alter their schedule, their approach -- make them second guess what they're doing and make sure they know that it's not going to be without some kind of cost''~\cite{nsa2022look}. These statements reflect a sludge-like strategy.

Nation-state cyber threats remain a focus for United States national security. ``We've got to put sand and friction in [adversary] operations so they don’t just get free shots on goal,'' said Rob Joyce, Director of the NSA's Cybersecurity Directorate, in 2021~\cite{aspen2021}. Persistent engagement, he explained further, is about more than offensive cyber; releasing information about tools and infrastructure is also successful in slowing adversaries.

\subsection{Election Security}
The protection of electoral systems against interference and influence is integral to democracy. In the United States this outcome involves the combined efforts of federal, state, local, and private sector partners. Election-related cyber threats have been observed in various forms from misinformation to denial of service attacks.

General Paul Nakasone, head of U.S. Cyber Command, testified before Congress ahead of the 2020 elections that ``USCYBERCOM is working with the combatant commands, DHS, FBI, across the Intelligence Community, and in conjunction with private sector and foreign partners to improve understanding and act to contest and frustrate adversary cyber activities''~\cite{nakasone2019statement}. Frustration is one byproduct of sludge when friction impedes an adversary's goal.

Election security is not limited to the United States. In the lead up to the 2017 French presidential election, Emmanuel Macron's campaign team deliberately created false email accounts and fake documents~\cite{nossiter2017hackers}. The stated goal of this deception was to slow down Russian attackers. When 
gigabytes of stolen data from  Macron's campaign were released online, it included real and forged emails. Despite speculation in the press about the effectiveness of the deception on the Russians, no official analysis was released.

\subsection{Russia-Ukraine Crisis}
The United States, together with dozens of other countries, imposed numerous sanctions against Russia in response to their military attack against Ukraine in early 2022. U.S. officials have reported that the financial sanctions successfully imposed cost, both monetary and psychological, on slowing some Russian cyber attacks. ``We've definitively seen the criminal actors in Russia complain that the functions of sanctions and the distance of their ability to use credit cards and other payment methods to get Western infrastructure to run these [ransomware] attacks have become much more difficult,'' Rob Joyce told The Cipher Brief~\cite{cipher2022view}.

USCYBERCOM, in partnership with the Security Service of Ukraine, also revealed malware used against Ukraine in order to disrupt cyber attacks~\cite{cnmf2022discloses}. This form of outing allows increased detection of malicious activity and thus imposes friction on the attackers who otherwise benefit from being undetected. While the release did not make a public attribution of the threat actor, it still may have slowed and disrupted the actors' activity.

\subsection{Ransomware}

Ransomware has become a serious and elusive cyber threat. Ransomware attacks rose during the COVID-19 pandemic when victims included healthcare, financial services, and government systems. Researchers and practitioners have proposed various technical countermeasures, including ransomware honeypots and honeyfiles~\cite{beaman2021ransomware}. Nevertheless, attacks remain persistent.

The challenge of ransomware is not simply technical, but also because of the ability for criminals to profit from it. Safe harbor in some nation-states limits the ability for criminal prosecution. However, international financial transactions are essential for ransomware payments and the more friction to financial benefits, the higher the opportunity cost for attackers.  

Rob Joyce reported in May 2022 that ransomware activity had declined in the early months of 2022, a trend he attributed, in part, to sanctions against Russia making it more difficult for attackers to procure infrastructure and transfer money~\cite{cyberuk2022plenary}. Thus, sludge had a desirable effect on slowing ransomware.

\section{Limitations and Future Work}

There are several considerations to the Sludge Strategy. First, we acknowledge that despite point solutions such as honeypots and emerging government examples, there are nascent implementations and evaluations of sludge as a strategy to impose cost on attackers. We hope that our work encourages further exploration and evaluation. Second, the effectiveness of the strategy may benefit from knowledge of individual human and organizational factors of the attackers. Sludge is, after all, an effect against the human nature of the attacker. If cyber adversaries embody common personalities and traits, then these would enable many kinds of sludge to be effective regardless of individual differences. Still, future research should examine the traits and characteristics of people for whom various sludge is effective.
Third, sludge will not stop a dedicated attacker who has enough resources to persist and adapt despite sludge, even though sludge will still slow them. Sludge implementations will have to respond and evolve to this type of adversary and also as general cybersecurity evolves. Nevertheless, persistent threat actors are still human and vulnerable to human weaknesses and psychological manipulation.

Cybersecurity professionals often seek to minimize their recovery time, failure rates, and lead times. If adversaries behave likewise, sludge may be used to strategically maximize negative results. That is, it may be possible to slow an adversary's recovery time or increase failure rates in capability development or operations. For example, according to a report, ``[Stuxnet] generated malfunctions in the centrifuges of the Natanz enrichment plant at random intervals over months, using different errors every time, and rendering them undetectable to the diagnostic systems in the control room''~\cite{greenbert2012stuxnet}.

Future work should explore new types of sludge. One idea is the potential value of fracturing adversary teams from within by employing nudges and sludge. Organized attackers commonly rely on teams with specialized roles.
Each human element in attack operations is subject to cognitive bias and the natural reticence to admit fault, leading to pointing fingers at each other when things go wrong. This internal division makes it harder to conduct successful operations. Sludge could enable defenders to slow down individual parts of this process.

The Sludge Strategy introduces new cost-imposing cyber defense by strategically deploying friction for attackers. The strategy broadens the options beyond complete denial and is consistent with modern information defense which assumes system and network compromise. We encourage cyber defenders, military planners, and system designers to consider how sludge can improve cybersecurity in their environments. New implementations of sludge must be developed and evaluated by interdisciplinary teams to ensure that they produce the desired outcomes.

\newpage
\bibliographystyle{IEEEtran}
\bibliography{bibliography}

\end{document}